\documentclass[aps,prl,twocolumn,english,showpacs,superscriptaddress,amssymb,amsfonts]{revtex4}
\usepackage[T1]{fontenc}
\usepackage[latin9]{inputenc}
\usepackage{amsmath}
\usepackage{dsfont}
\usepackage{amstext}

\usepackage{tocvsec2}

\usepackage{amssymb}
\usepackage{amsbsy}
\usepackage{amsthm}
\usepackage{epsfig}
\usepackage{framed}
\usepackage{graphicx}
\usepackage{bbm}
\usepackage{hyperref}
\usepackage{color}
\usepackage{multirow}

\newcommand{\Ref}[1]{Ref.~\onlinecite{#1}}

\newcommand{\bst}{{\mathcal{T}}}

\newcommand{\ie}{{\emph{i.e.~}}}
\makeatletter

\newcommand{\Rmnum}[1]{\expandafter\@slowromancap\romannumeral #1@}
\makeatother
\newcommand{\imth}{\hspace{1pt}\mathrm{i}\hspace{1pt}}

\newcommand{\eg}{{\emph{e.g.~}}}

\newcommand{\mbz}{{\mathbb{Z}}}
\newcommand{\bea}{\begin{eqnarray}}
\newcommand{\eea}{\end{eqnarray}}
\newcommand{\bpm}{\begin{pmatrix}}
\newcommand{\epm}{\end{pmatrix}}
\newcommand{\bal}{\begin{aligned}}
\newcommand{\eal}{\end{aligned}}

\newcommand{\expval}[1]{\langle{#1}\rangle}

\newcommand{\dket}[1]{|{#1}\rangle}
\newcommand{\dbra}[1]{\langle{#1}|}

\newtheorem{theorem}{Theorem}

\makeatother

\usepackage{babel}
\begin{document}
\title{Lieb-Schultz-Mattis theorems for symmetry-protected topological phases}

\author{Yuan-Ming Lu}
\affiliation{Department of Physics, The Ohio State University, Columbus, OH 43210, USA}

\begin{abstract}
The Lieb-Schultz-Mattis (LSM) theorem and its descendants represent a class of powerful no-go theorems that rule out any short-range-entangled (SRE) symmetric ground state irrespective of the specific Hamiltonian, based only on certain microscopic inputs such as symmetries and particle filling numbers. In this work, we introduce and prove a new class of LSM-type theorems, where any symmetry-allowed SRE ground state must be a symmetry-protected topological (SPT) phase with robust gapless edge states. The key ingredient is to replace the lattice translation symmetry in usual LSM theorems by magnetic translation symmetry. These theorems provide new insights into numerical models and experimental realizations of SPT phases in interacting bosons and fermions.
\end{abstract}


\maketitle





The Lieb-Schultz-Mattis (LSM) theorem\cite{Lieb1961} and its descendants\cite{Oshikawa2000,Hastings2004,Hastings2005,Parameswaran2013,Watanabe2015,Po2017} are powerful theorems that dictate long-distance low-energy (infrared) properties of a lattice-translation-invariant system from its microscopic (ultraviolet) input, such as a global $U(1)$ charge/spin conservation symmetry and the filling number per unit cell (u.c.). Irrespective of the microscopic Hamiltonian, remarkably, these generic ultraviolet inputs dictate that a lattice-translation-invariant ground state at a non-integer filling is either gapless (\eg in metals), or spontaneous breaks the $U(1)$ symmetry (\eg in superconductors), or develops intrinsic topological orders\cite{Wen2004B} (\eg in fractional quantum Hall states). In all cases, the system forbids a short-range-entangled (SRE) ground state\cite{Zeng2015} that preserves both global and lattice translation symmetries (\eg a featureless Mott insulator without fractionalization). Since LSM theorems apply to a generic interacting system, they provide great insights in the study of quantum many-body systems beyond one spatial dimension (1d), which remains intractable in most analytic or numeric efforts.

While usual LSM theorems forbids a SRE symmetric ground state at fractional fillings, the interplay of symmetry and topology gives rise to a rich structure of SRE symmetric states, coined symmetry protected topological (SPT) phases\cite{Chen2013,Senthil2014}. Characterized by protected edge/surface states, topological insulators and superconductors\cite{Hasan2010,Qi2011} are the examples of SPT phases in non-interacting fermions. In spite of extensive theoretical studies, so far strongly-interacting SPT phases are still in lack of realizations beyond 1d spin chains\cite{Haldane1983a,Affleck1987}.

In this work we intend to fill this gap by introducing and proving a new class of LSM theorems in TABLE \ref{tab:2d:fermion}-\ref{tab:2d:boson}, whose SRE symmetric ground state must be a SPT phase. Focusing on two spatial dimensions (2d), we show that the key step is to replace the pure lattice translations in usual LSM theorem by magnetic translations\cite{Zak1964a}, with a rational $\phi=2\pi\frac pq$ flux piercing through each plaquette. We will always refer a unit cell (u.c.) as the original cell generated by Bravais lattice translations, in contrast to enlarged magnetic unit cell containing one flux quantum (or $2\pi$ flux). We show that a symmetric SRE ground state with proper degrees of freedom (d.o.f.) per u.c. (such as fractional filling $\bar\rho\notin\mbz$) must be a SPT phase with protected edge modes. As will become clear later, a key idea is the charge-flux binding in SPT phases\cite{Chen2014}.

\section{LSM theorems for fermion SPT phases}

Topological insulators (TIs) and topological superconductors (TSCs) are well-known examples of fermion SPT phases. In the Altland-Zirnbauer (AZ) 10-fold way of symmetry classes, nontrivial fermion SPT phases exist in 5 symmetry classes in each spatial dimension\cite{Schnyder2008,Kitaev2009}. In two spatial dimensions (2d), below we demonstrate that a LSM-type theorem favoring a SPT ground state exists for 4 AZ symmetry classes: \ie class D, DIII, A and AII as summarized in TABLE \ref{tab:2d:fermion}.

{\bf Symmetry class D} describes superconductors with no symmetry, with a $\mbz$ classification in 2d. Characterized by an integer-valued topological index $\nu\in\mbz$, they host chiral Majorana edge modes with a chiral central charge $c_-=\nu/2$\cite{Kitaev2006}. The simplest $\nu=1$ SPT phase is the chiral $p_x+\imth p_y$ superconductor of spinless (or spin-polarized) electrons in 2d. One significant property of a $\nu=$~odd topological superconductor is an odd number of Majorana zero modes (MZMs) localized at each superconducting vortex core\cite{Read2000}, robust against any perturbations. One MZM can be viewed as ``half'' of a fermion and it has been proven that a unique symmetric SRE ground state is not allowed in a translational invariant system with odd Majoranas per unit cell (u.c.)\cite{Hsieh2016}. On the other hand, magnetic translations allow a SRE TSC ground state, as our theorem states:
\begin{theorem}
\label{thm:f:class D}
For a generic interacting fermion system with an odd number of Majoranas per u.c., in the presence of magnetic translation symmetry\cite{Zak1964a}
\bea\label{magnetic translation}
\tilde T_1\tilde T_2\tilde T_1^{-1}\tilde T_2^{-1}=e^{\imth\phi\hat F},~~~\hat F=\text{total fermion number}.
\eea
with $\phi=\pi$ flux per u.c., if there is a unique symmetric and gapped ground state on torus, it must be a $\nu=$~odd TSC in class D with chiral Majorana edge states.
\end{theorem}

Now that all $\nu=$~odd chiral TSC necessarily breaks time reversal symmetry, there is a no-go theorem as a straightforward corollary of theorem \ref{thm:f:class D}:

\emph{For a generic interacting fermion system with an odd number of Majoranas per u.c., in the presence of time reversal and magnetic translation symmetry (\ref{magnetic translation}) with $\phi=\pi$, it is impossible to have a symmetry-preserving unique gapped ground state on torus.}

\begin{figure}[h]
\centering
\includegraphics[width=0.4\columnwidth]{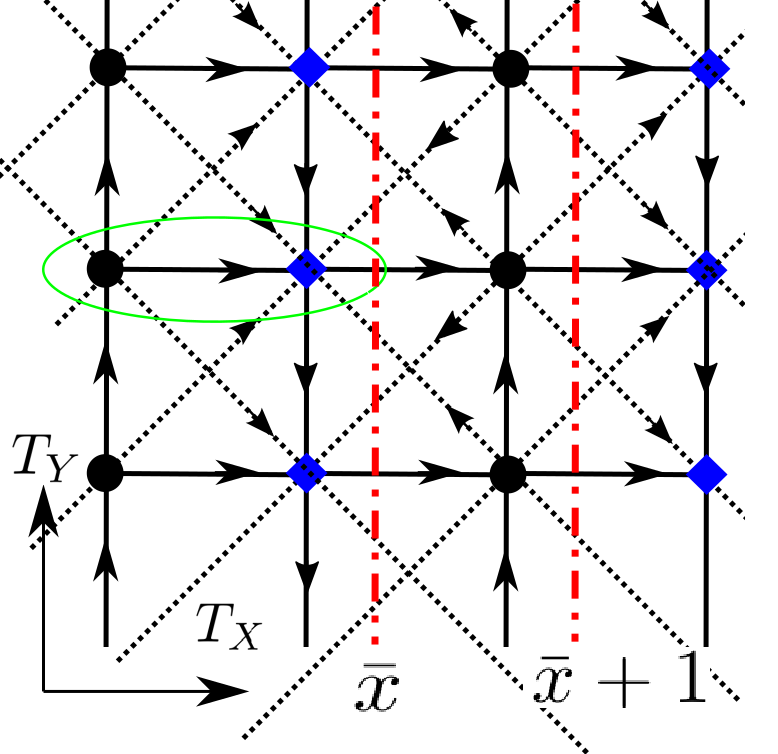}
\includegraphics[width=0.58\columnwidth]{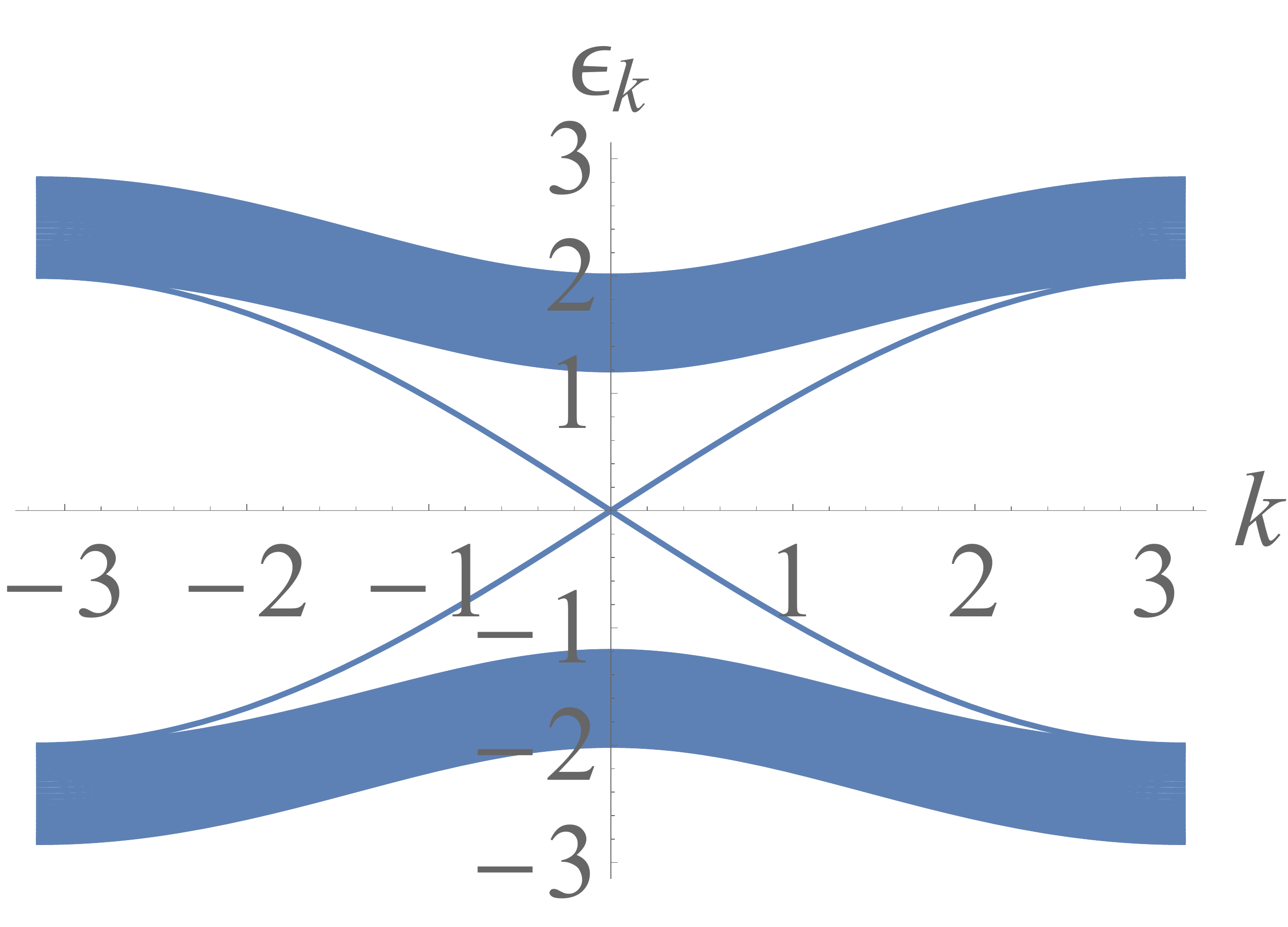}
\caption{(color online) Majorana hopping model (\ref{pi-flux:D}) with $\phi=\pi$ flux per u.c. on square lattice (left) and its edge spectrum (right). Arrows represent the signs of Majorana hoppings, while the green oval stands for the doubled magnetic u.c.. The edge spectrum is obtained on a $L_y=50$ open cylinder (periodic along $\hat x$ direction), where NNN coupling is chosen as $t_2/t_1=0.3$ with NN couplings $t_x=t_y=t_1$. The two edge modes with opposite chirality are located on two open edges separately.}
\label{fig:square}
\end{figure}

We demonstrate Theorem \ref{thm:f:class D} by a square lattice $\pi$-flux model with 1 Majorana $\gamma_{\bf r}$ at each site ${\bf r}=(x,y)$, as illustrated in FIG. \ref{fig:square}. Choosing a Landau gauge for magnetic translation algebra (\ref{magnetic translation})
\bea\label{landau gauge}
\tilde T_y=T_y,~~~\tilde T_x=T_x e^{\imth\phi\sum_{\bf r}y\hat F_{\bf r}}
\eea
where $\hat F_{\bf r}$ is the fermion number on site ${\bf r}$, the nearest neighbor (NN) $\phi=\pi$-flux Hamiltonian writes
\bea\label{pi-flux:D}
\hat H_0^{\text D}=\sum_{\bf r}\imth[t_x\gamma_{\bf r}\gamma_{{\bf r}+\hat x}+t_y(-1)^x\gamma_{\bf r}\gamma_{{\bf r}+\hat y}]+h.c.
\eea
It's straightforward to identify two zero-energy Majorana cones at $(k_x,k_y)=(0,0)$ and $(0,\pi)$ related by magnetic translation $\tilde T_x$ in (\ref{landau gauge}), and there is only one mass term $m\hat \Gamma_0$ allowed by $\tilde T_{x,y}$ symmetries\cite{Supp}, realized by \eg next nearest-neighbor (NNN) hoppings between Majoranas (see FIG. \ref{fig:square}). In the usual band inversion story of TIs, opposite signs of mass $m$ lead to a trivial insulator and a TI respectively. However for the Majorana cones in model (\ref{pi-flux:D}), both signs of the mass term lead to a TSC in class D, whose topological index $\nu=\text{Sgn}(m)$ and chirality of edge modes depends on the mass sign. In FIG. \ref{fig:square} we show the chiral edge modes of such a symmetry-enforced TSC on square lattice.

{\bf Symmetry class DIII} describes time-reversal-invariant (TRI) superconductors, with a $\mbz_2$ classification in 2d. The TSC in class DIII is a triplet TRI $p$-wave superconductor\cite{Qi2009}, a 2d analog of $^3$He B phase. One of its defining character is one Kramers pair of MZMs $\{\gamma_{\uparrow},\gamma_{\downarrow}\}$ at each vortex core, stable against any time-reversal-invariant perturbations. One can also prove a no-go theorem\cite{Supp} that rules out any symmetric SRE ground state in a translational invariant system with an odd number of Majorana Kramers pairs $\{\gamma_{a,\uparrow},\gamma_{a,\downarrow}|1\leq a<2N\}$ per u.c.. On the other hand, similar to class D, magnetic translation symmetry however allows a SRE TSC ground state:
\begin{theorem}
\label{thm:f:class DIII}
For a generic interacting fermion system with {an odd number of Majorana Kramers pairs per u.c.}, in the presence of time reversal symmetry $\hat\bst^2=(-1)^{\hat F}$ and magnetic translation symmetry (\ref{magnetic translation}) with $\phi=\pi$ flux per u.c., any unique symmetric and gapped ground state on torus must be a TSC in class DIII with helical Majorana edge states.
\end{theorem}

To demonstrate Theorem \ref{thm:f:class DIII}, we again consider a square lattice NN $\pi$-flux model with one Kramers pair $\{\gamma_{{\bf r},\uparrow},\gamma_{{\bf r},\downarrow}\}$ of Majoranas per site ${\bf r}$:
\bea\label{pi-flux:DIII}
\hat H_0^{\text{DIII}}=\sum_{{\bf r},\sigma}\imth\sigma[t_x\gamma_{{\bf r},\sigma}\gamma_{{{\bf r}+\hat x},\sigma}+t_y(-1)^x\gamma_{{\bf r},\sigma}\gamma_{{\bf r}+\hat y,\sigma}]+h.c.
\eea
Similar to model (\ref{pi-flux:D}) in class D, in the basis of $\phi_{\bf k}=\frac1{\sqrt{L_xL_y/2}}\sum_{(x,y)}e^{-\imth(k_xx+k_yy)}(\gamma_{(2x,y),\sigma},\gamma_{(2x+1,y),\sigma})^T$, it's straightforward to show that NN model (\ref{pi-flux:DIII}) leads to two Dirac points at ${\bf k}=(0,0)$ and $(0,\pi)$, described by low-energy Dirac Hamiltonian
\bea\label{dirac ham:class DIII}
\hat H_0^\text{DIII}\rightarrow-\sum_{|{\bf q}|\ll1}\Phi^\dagger_{\bf q}(\frac{t_x}2q_x\tau_x+t_yq_y\tau_z\mu_z)\sigma_z\Phi_{\bf q}+O(|{\bf q}|^2),
\eea
where $\vec\tau$, $\vec\mu$ and $\vec\sigma$ are Pauli matrices for sublattice (in a doubled magnetic cell), valley and spin indices. With the following symmetry operations on Dirac spinor $\Phi_{\bf q}$:
\bea
\Phi_{\bf q}\overset{T_y}\longrightarrow\mu_z\Phi_{\bf q},~~\Phi_{\bf q}\overset{\tilde T_x}\longrightarrow\tau_x\mu_x\Phi_{\bf q},~~\Phi_{\bf q}\overset{\bst}\longrightarrow\imth\sigma_y\Phi_{-\bf q}.\notag
\eea
There is only one symmetric mass term for Dirac Hamiltonian (\ref{dirac ham:class DIII}) $\hat\Gamma_0=\tau_y\mu_z\sigma_z$, realized by \eg TRI NNN couplings shown in FIG. \ref{fig:square}. Irrespective of the mass sign, the gapped Dirac Hamiltonian always leads to a TRI TSC with helical Majorana edge modes.

\begin{table*}[tb]
\centering
\begin{tabular} {|c|c||c|c|c||c|c|c|}
\hline
\multicolumn{2}{|c||}{Physical systems}&\multicolumn{3}{|c||}{Microscopic input}&\multicolumn{3}{|c|}{Output of LSM theorem}\\
\hline
\multirow{2}{1cm}{AZ\cite{Altland1997} class}&\multirow{2}{2cm}{Applications}&\multirow{2}{1.5cm}{Symmetry group} &
\multirow{2}{1.4cm}{d.o.f. per unit cell}&\multirow{2}{1.5cm}{Flux per unit cell}&\multirow{2}{1cm}{Topological invariant}&{Edge states}&\multirow{2}{2cm}{Chiral central charge}\\
&&&&&&&\\
\hline
D&Majorana&$(-1)^{\hat F}$&\multirow{2}{2.5cm}{Odd Majoranas $\{\gamma_1,\cdots,\gamma_{2N-1}\}$}&$\phi=\pi$&$\nu=$~odd&Chiral Majorana&$c_-=\nu/2$\\
&vortex lattice\cite{Grosfeld2006}&&&&$\in\mbz$&&\\ \hline
DIII&\multirow{2}{2cm}{Kitaev-type spin liquid\cite{Nakai2012}}&$\hat\bst^2=(-1)^{\hat F}$&\multirow{2}{2.5cm}{Odd Majorana Kramers pairs} 
&$\phi=\pi$&$\nu=1$&Helical Majorana&$c_-=0$\\
&&&&&$\in\mbz_2=\{0,1\}$&&\\
\hline
A&\multirow{2}{2.5cm}{Integer QHE in Hofstadter model}&$U(1)_{\text{charge}}$&Charge $e\cdot\bar\rho_f$&$\phi=2\pi\frac pq$&$p\sigma_{xy}=q\bar\rho_f\mod q$&Chiral fermion&$c_-=\sigma_{xy}$\\
&&&&&$\sigma_{xy}\in\mbz$&&\\
\hline
AII&\multirow{2}{2.2cm}{QSHE in $\pi$-flux model\cite{Wu2017}}&\multirow{2}{1.8cm}{$U(1)_{\text{charge}}$ $\hat\bst^2=(-1)^{\hat F}$}&Charge $e\cdot \bar\rho_f$&$\phi=\pi$&$\nu=1$&Helical fermion&$c_-=0$\\
&&&($\bar\rho_f=$~odd)&&$\in\mbz_2=\{0,1\}$&&\\
\hline
\end{tabular}
\caption{Summary of LSM theorems for SPT phases (\ie topological insulators/superconductors) of a generic interacting fermion system in two dimensions. The trivial insulators/superconductors have topological index $\nu,\sigma_{xy}=0$.}
\label{tab:2d:fermion}
\end{table*}

{\bf Symmetry class A} describes fermion insulators (or metals) with $U(1)$ charge conservation symmetry, where lattice translation symmetry allows a well-defined filling number $\bar\rho_f$ per u.c.. The SPT phases in class A corresponds to Chern insulators characterized by integer-valued Hall conductance $\sigma_{xy}\in\mbz$ (in unit of $e^2/h$)\cite{Thouless1982}. In the usual LSM theorem\cite{Lieb1961,Oshikawa2000,Hastings2004} with lattice translations, an insulating ground state without fractionalization is impossible at non-integer fillings $\bar\rho_f\notin\mbz$. However with $\phi$ flux per u.c. and associated magnetic translation symmetry (\ref{magnetic translation}), an unfractionalized SPT ground state becomes a possibility even at a fractional filling\cite{Lu2017}. Specifically the Hall conductance $\sigma_{xy}$ is constrained by the following LSM theorem\cite{Lu2017}:
\begin{theorem}
\label{thm:f:class A}
Consider a generic interacting fermion system preserving $U(1)$ charge conservation and magnetic translation symmetry (\ref{magnetic translation}), with charge density $\bar\rho$ and flux density $\phi$ per u.c., if there is a unique symmetric and gapped ground state on torus, its Hall conductance $\sigma_{xy}$ satisfy the following condition:
\bea\label{lsm:U(1)}
\sigma_{xy}\cdot\frac{\phi}{2\pi}=\bar\rho\mod1.
\eea
\end{theorem}
In a special case, half-filled ($\bar\rho_f=1/2$) fermions with $\phi=\pi$ must have an odd Hall conductance $\sigma_{xy}=1\mod2$ for any unique gapped ground state, therefore forbidding a unique gapped TRI ground state. This is consistent with the well-known Dirac spectrum of NN $\pi$-flux model on square lattice\cite{Affleck1988a}.

{\bf Symmetry class AII} describes half-integer-spin fermions preserving $U(1)$ charge conservation and time reversal symmetry $\bst$ with $\bst^2=(-1)^{\hat F}$. The SRE fermion phases are characterized by a $\mbz_2$-valued index $\nu=0,1$ where $\nu=1$ corresponds to a QSH insulator with protected helical edge modes. With lattice translation symmetry, an unfractionalized insulating ground state is only possible at even fermion filling $\bar\rho_f\in2\mbz$\cite{Watanabe2015}. Time reversal symmetry is only compatible with $\phi=\pi$ flux per u.c., and the associated magnetic translation symmetry (\ref{magnetic translation}) brings in a new possibility at an odd filling\cite{Wu2017}:
\begin{theorem}
\label{thm:f:class AII}
Consider a generic interacting fermion system preserving $U(1)$ charge conservation, time reversal $\bst^2=(-1)^{\hat F}$ and magnetic translation symmetry (\ref{magnetic translation}), with fermion density $\bar\rho_f=1\mod2$ and $\phi=\pi$ flux per u.c., if there is a unique symmetric and gapped ground state on torus, it must be a QSH insulator.
\end{theorem}

\begin{table*}[tb]
\centering
\begin{tabular} {|c|c|c||c|c||c|}
\hline
\multicolumn{3}{|c||}{Global symmetries and SPT classification}&\multicolumn{2}{|c||}{Microscopic input}&{Output of LSM theorem}\\
\hline
\multirow{2}{1.5cm}{Symmetry group $G_s$} &\multirow{2}{1.9cm}{Classification $\mathcal{H}^3(G_s,U(1))$}&Topological invariants&
\multirow{2}{2cm}{Density/d.o.f. per unit cell}&\multirow{2}{1.5cm}{Flux per unit cell}&{Topological index}\\
&&&&&\\
\hline
$U(1)$&$2\mbz$&$\sigma_{xy}=$~even&$\bar\rho=\frac{2a}{q}$&$\phi=2\pi\frac{p}{q}$&$p\cdot\sigma_{xy}=2a\mod q$\\
&&&&&(BIQH states)\\
\hline
$U(1)\rtimes Z_2^\bst$&$\mbz_2$&$\nu=0,1\in\mbz_2$&$\bar\rho\in\mbz$ + an odd number&$\phi=\pi$&$\nu=1$\\
&&&of Kramers doublets&&(BQSH state)\\
\hline
$Z_2\times Z_2^\bst$&$\mbz_2\times\mbz_2$&$\nu,\nu_\bst=0,1\in\mbz_2$&an odd number&$\phi=\pi$&$\nu_\bst=1$\\
&&&of Kramers doublets&&\\
\hline
$U(1)_A\times U(1)_B$&$(2\mbz)^2\times\mbz$&\multirow{2}{2.3cm}{$\sigma_{xy}^{A},\sigma_{xy}^B=$~{even} $\sigma_{xy}^{AB}=\sigma_{xy}^{BA}\in\mbz$}&$(\bar\rho_A,\bar\rho_B)$&$(\phi_A,\phi_B)$&$(\sigma^{A}_{xy},\sigma_{xy}^B,\sigma_{xy}^{AB})$ satisfying (\ref{boson:U(1)xU(1)})\\
&&&&&\\
\hline
$U(1)_A\times (Z_q)_B$&$2\mbz\times (\mbz_q)^2$&\multirow{2}{2.3cm}{$\sigma_{xy}^A=$~{even} $\nu^{B},\nu^{AB}\in\mbz_q$}&$\bar\rho_A=\frac{a}{q}$&\multirow{2}{1.5cm}{$\phi_B=2\pi\frac{p}{q}$ $\phi_A=0,\pi$}&$p\cdot \nu^{AB}=a\mod q$ \\
&&&&&\\
\hline
\end{tabular}
\caption{Summary of LSM theorems for SPT phases of a generic interacting boson system in two dimensions. Hall conductance $\sigma_{xy}$ is defined in unit of $1/h$, where the unit charge of microscopic bosons is set to 1, $(p,q)$ are mutually-primed integers and $a\in\mbz$.}
\label{tab:2d:boson}
\end{table*}

\section{LSM theorems for boson SPT phases}

While TIs and TSCs are realizable even in a system of non-interacting (free) fermions, in a boson system strong interactions are necessary to evade Bose-Einstein condensation and to achieve a gapped symmetric ground state. Among them, boson SPT phases with symmetry group $G_s$ are SRE symmetric ground states with $G_s$-symmetry-protected edge/surface excitations. Below we present LSM theorems for various global symmetry $G_s$ (see TABLE \ref{tab:2d:boson}) containing $U(1)\subset G_s$ as a subgroup. We label the $U(1)$ charge density per u.c. as $\bar\rho$.

One minor (notational) difference from fermions is that magnetic translation algebra (\ref{magnetic translation}) will be written in a more generic context:
\bea\label{magnetic translation:general}
\tilde T_1\tilde T_2\tilde T_1^{-1}\tilde T_2^{-1}=e^{\imth\phi\hat N},~~~\hat N=\text{total}~U(1)~\text{charge}.
\eea
where $\phi$ is the flux per u.c. associated with the $U(1)$ symmetry. Similar to the TSC case, $U(1)$ charge conservation is not required to define the above magnetic translation symmetry. Even if the $U(1)$ group is broken down to a discrete $Z_q$ subgroup generated by $\hat R_q\equiv e^{\imth\frac{2\pi}{q}\hat N}$, a flux of $\phi=\frac{2\pi p}{q}$ (with $p,q\in\mbz$) is still well-defined in (\ref{magnetic translation:general}).

{\bf Boson integer quantum Hall (QIHE) states}: In a simplest case we consider a boson system with $G_s=U(1)$. The associated boson SPT phases are BIQH states, characterized by an even Hall conductance $\sigma_{xy}\in2\mbz$ (in unti of $1/h$ where unit charge is set to 1). Analogous to symmetry class A of fermions, the usual LSM theorem forbids a unique gapped ground state at any non-integer filling $\bar\rho\notin\mbz$ with lattice translation symmetry. In the presence of magnetic translation (\ref{magnetic translation:general}), a SPT ground state with $\sigma_{xy}\neq0$ becomes possible even at a fractional filling. Specifically, the LSM theorem for interacting bosons with $G_s=U(1)$ symmetry has the same form as Theorem \ref{thm:f:class A} for fermions, also yielding the constraint (\ref{lsm:U(1)}).

For any rational flux density $\phi=2\pi\frac pq$ and commensurate charge density $\bar\rho=\frac{2a}q$, we have $p\sigma_{xy}=2a\mod q$ for any SRE symmetric ground state shown in TABLE \ref{tab:2d:boson}. This necessarily leads to a nonzero Hall conductance, thus a BIQH state.

{\bf Bosonic quantum spin Hall (BQSH) states}: With both $U(1)$ charge conservation and time reversal symmetry \ie $G_s=U(1)\rtimes Z_2^\bst$, SRE symmetric boson states are classified by a $\mbz_2$-valued index $\nu=0,1$, where $\nu=1$ corresponds to the nontrivial BQSH state\cite{Lu2012a}. In addition to protected edge states\cite{Lu2012a} there is another defining character for BQSH states: each $\pi$ flux in the bulk is bound to a Kramers doublet transformed as $\bst^2=-1$\cite{Chen2014}. A half-integer spin is an example of a Kramers doublet in contrast to an integer spin. If there is an odd number of Kramers doublets per u.c., usual LSM theorems with lattice translation symmetry forbids any SRE symmetric ground state \cite{Watanabe2015}. In the presence of magnetic translation (\ref{magnetic translation:general}) with TRI $\phi=\pi$ flux per u.c., however, a BQSH ground state with an odd number of Kramers doublets per u.c. becomes possible:
\begin{theorem}
\label{thm:b:U(1)+T}
Consider a generic interacting boson system preserving $U(1)$ charge conservation, time reversal $\bst$ and magnetic translation symmetry (\ref{magnetic translation:general}), with charge density $\bar\rho\in\mbz$, flux density $\phi=\pi$ and an odd number of Kramers doublets (e.g. spin-$1/2$'s with $\bst^2=-1$) per u.c., if there is a unique symmetric and gapped ground state on torus, it must be a BQSH state.
\end{theorem}

The boson system in Theorem \ref{thm:b:U(1)+T} consists of two parts: charged bosons that transform as Kramers singlets ($\bst^2=+1$), and half-integer spins that transform as Kramers doublets ($\bst^2=-1$). Therefore the $\pi$ flux per u.c. is only visible to charge d.o.f. but not to spins. In fact, the $U(1)$ symmetry above can be broken down to a discrete $Z_2$ subgroup, resulting in a symmetry group $G_s=Z_2\times Z_2^\bst$. The associated SRE symmetric states have a $(\mbz_2)^2$ classification\cite{Chen2013,Lu2012a}, where one $\mbz_2$ index ($\nu_\bst$ in TABLE \ref{tab:2d:boson}) comes from the binding of each $\pi$ flux to a Kramers doublet\cite{Chen2014}. Although the charge will no longer conserve with $G_s=Z_2\times Z_2^\bst$, the $\pi$ flux per u.c. is still well-defined and our LSM states that \emph{any unique gapped ground state on torus with an odd number of Kramers doublets per u.c. must be a SPT state} where a $\pi$ flux is bound to a Kramers doublet\cite{Supp}.

{\bf Two-component BIQH states}: For a two-component system with two species of conserved bosons, symmetry group $G_s=U(1)_A\times U(1)_B$ leads to $(2\mbz)^2\times\mbz$ classification of SRE symmetric states. They are characterized by a (real symmetric) $2\times2$ Hall conductance tensor $\sigma_{xy}^{\alpha,\beta},~\alpha,\beta=A,B$. While the intra-species Hall conductance $\sigma_{xy}^{A},\sigma_{xy}^{B}\in2\mbz$ must be even integers, the inter-species Hall conductance $\sigma_{xy}^{AB}=\sigma_{xy}^{BA}\in\mbz$ can take any integer value. The magnetic translation symmetry here is defined by two flux $\phi_A$ and $\phi_B$ for the two components:
\bea\label{magnetic translation:two}
\tilde T_1\tilde T_2\tilde T_1^{-1}\tilde T_2^{-1}=e^{\imth(\phi_A\hat N_A+\phi_B\hat N_B)}.
\eea
Our LSM theorem states the following:
\begin{theorem}
\label{thm:b:U(1)xU(1)}
Consider a generic interacting boson system of two components A and B, separately conserved with a symmetry group $G_s=U(1)_A\times U(1)_B$. In the presence of magnetic translation symmetry (\ref{magnetic translation:two}), with charge density $(\bar\rho_A,\bar\rho_B)$ and flux density $(\phi_A,\phi_B)$ per u.c., if there is a unique symmetric and gapped ground state on torus, its Hall conductance tensor must satisfy
\bea\label{boson:U(1)xU(1)}
\frac1{2\pi}\bpm\sigma_{xy}^A&\sigma_{xy}^{AB}\\ \sigma_{xy}^{BA}&\sigma_{xy}^B\epm\bpm\phi_A\\ \phi_B\epm=\bpm\bar\rho_A\\ \bar\rho_B\epm\mod1
\eea
\end{theorem}
This can be viewed as a generalization of Theorem (\ref{thm:f:class A}) with $G_s=U(1)$, and can be further generalized to a multi-component system. At any fractional filling \ie $\rho_{A,B}\notin\mbz$, the Hall conductance tensor cannot vanish identically, leading to a SPT ground state. Furthermore, one of the two $U(1)$ symmetries here can be broken down to a $Z_q$ subgroup, as we discuss below.

{\bf Two-component magnets}: In a two-component magnetic system where species $A$ has a $U(1)$ conservation (\eg of $\hat z$-component spin) and species $B$ with only a discrete $Z_q$ symmetry, the global symmetry $G_s=U(1)_A\times(Z_q)_B$ leads to a $2\mbz\times(\mbz_q)^2$ classification of 2d SPT phases\cite{Chen2013}. Compared to previous $G_s=U(1)_A\times U(1)_B$ case, while $\sigma_{xy}^{A}\in2\mbz$ still serves as a valid topological index with $U(1)_A$ symmetry, $\sigma_{xy}^{AB}=\sigma_{xy}^{BA}$ and $\sigma_{xy}^{B}$ are only well-defined modular $q$ once $U(1)_B$ is broken down to $(Z_q)_B$, yield two $\mbz_q$-valued indices:
\bea
\nu^{AB}\equiv\sigma_{xy}^{AB}=\sigma_{xy}^{BA}\mod q,~~~\nu^B\equiv\sigma_{xy}^{B}\mod2q.\notag
\eea
Our LSM theorem for this case reads the following:
\begin{theorem}
\label{thm:b:U(1)xZq}
Consider a generic interacting two-component (A and B) spin system with global symmetry $G_s=U(1)_A\times(Z_q)_B$, in the presence of magnetic translation symmetry (\ref{magnetic translation:two}), with conserved $U(1)_A$ density $\bar\rho_A$ and flux density $(\phi_A,\phi_B=2\pi\frac pq)$ per u.c., if there is a unique symmetric and gapped ground state on torus, its topological indices must satisfy
\bea\label{boson:U(1)xZq}
\sigma_{xy}^A\frac{\phi_A}{2\pi}+\nu^{AB}\frac{\phi_B}{2\pi}=\sigma_{xy}^A\frac{\phi_A}{2\pi}+\nu^{AB}\frac pq=\bar\rho_A\mod1.~~
\eea
\end{theorem}

Now that $\sigma_{xy}^A$ is an even integer, by choosing flux $\phi_A=0,\pi$ condition (\ref{boson:U(1)xZq}) immediately leads to $p\cdot\nu^{AB}=q\bar\rho_A\mod q$ as shown in TABLE \ref{tab:2d:boson}. This indicates any SRE symmetric ground state at fractional filling $\bar\rho_A\notin\mbz$ must be a SPT state with $\nu^{AB}\neq0\mod q$.

\section{Physical picture and sketch of proof}

What are the basic ideas behind these LSM theorems for SPT phases? Given the global and lattice translation symmetries, the usual LSM theorems dictate the ``integer filling'' (per u.c.) conditions on the Hilbert space that allow a symmetric SRE ground state: \eg integer filling $\bar\rho$ per u.c. for conserved $U(1)$ charges\cite{Lieb1961,Oshikawa2000,Hastings2004}, an even number of Kramers doublets per u.c. for TRI systems\cite{Watanabe2015}, and an even number of Majorana fermions per u.c. for superconductors\cite{Hsieh2016}. When pure lattice translations are replaced by magnetic translations, there is an additional ``background flux'' in each u.c., in addition to ``bare'' symmetry charges (the ``filling number'').
A key feature of many SPT phases is the binding of a ``fractionalized'' symmetry charge to a symmetry flux\cite{Roy2010}: \eg charge-flux binding in quantum Hall states, the binding of a Kramers doublet to each $\pi$-flux in QSH states\cite{Ran2008a,Qi2008a}, and the binding of a MZM to each $\pi$ flux in a chiral TSC\cite{Read2000}. For SPT states with magnetic translation symmetries, a ``background polarization charge'' comes together with the background flux, leading to an total symmetry charge different from the ``bare'' symmetry charges. This explains why a SPT ground state at fractional filling evades the usual LSM theorem requiring integer fillings, with the help of magnetic translations.

Take $G_s=U(1)$ as an example, with bare symmetry charge $\bar\rho$ and background polarization charge $-\sigma_{xy}\frac{\phi}{2\pi}$, the ``total charge'' per u.c. $\rho_\text{total}=\bar\rho-\sigma_{xy}\frac{\phi}{2\pi}$ must be an integer as dictated by the usual LSM theorem. This is exactly the condition (\ref{lsm:U(1)}) in Theorem \ref{thm:f:class A}.

This simple physical picture not only provides a generic construction of these LSM theorems, but also leads to their proofs. Below we only sketch the proofs, leaving details to supplemental materials. Consider a many-body system on a $L_x\times L_y$ torus (or an infinitely-long cylinder, periodic along $\hat y$ direction with a finite circumference $L_y$). In the presence of magnetic translation symmetry (\ref{magnetic translation}) or (\ref{magnetic translation:general}) with $\phi$ flux per u.c., we choose a circumference $L_y$ satisfying $\phi L_y\neq0\mod2\pi$: \eg for $\phi=\pi$ flux we choose $L_y=$~odd. In the Landau gauge (\ref{landau gauge}) with pure lattice translation symmetry $\tilde T_y=T_y$ along $\hat y$ direction, clearly the boundary condition along $\hat y$ direction changes with the coordinate $x$ (hence no well-defined magnetic translation $\tilde T_x$). Specifically the $\hat y$-direction boundary condition $\Theta_y$ of a many-body state $\dket{\Psi(\Theta_y)}$ with wavefunctions $\Psi({\bf X}_1,{\bf X}_2,\cdots)$ can be defined as\cite{Haldane1985b} (${\bf X}_i\equiv(x_i,y_i)$ are coordinates of $i$-th particle)
\bea\label{bc:y}
&\Psi({\bf X}_1+L_y\hat y,{\bf X}_2,\cdots)=e^{\imth\Theta_y}\Psi({\bf X}_1,{\bf X}_2,\cdots).
\eea
Equivalently it can be written as $e^{\imth L_y\hat K_y^1}\dket{\Psi(\Theta_y)}=e^{\imth\Theta_y}\dket{\Psi(\Theta_y)}$, where $\hat{\bf K}^1$ is the momentum of one single particle.
Although magnetic translation $\tilde T_x$ is absent on the torus/cylinder of chosen size $L_y$, the change of boundary condition $\Theta_y$ upon pure lattice translation $x\overset{\hat T_x}\rightarrow x+1$ leads to an important condition:
\bea\label{twist bc}
\hat T_x\dket{\Psi(\Theta_y)}=\dket{\Psi(\Theta_y+\phi L_y)}.
\eea
which imposes a strong constraint on any gapped and symmetric ground state on this torus/cylinder.

Take $G_s=U(1)$ (Thm. \ref{thm:f:class A}) for example, a SRE insulator can be characterized by a well-defined polarization\cite{Vanderbilt1993,RestaJPCMrev2002} $\hat P_x\equiv e^{\frac{2\pi}{L_x}\sum_{\bf r}x\hat n_{\bf r}}$ where $\hat n_{\bf r}$ labels the $U(1)$ charge on lattice site ${\bf r}$. Due to the non-commutative algebraic relation $\hat T_x\hat P_x\hat T_x^{-1}=e^{-\imth L_y\bar\rho}\hat P_x$, condition (\ref{twist bc}) implies the change of ground state polarization $\expval{\hat P_x}$ upon twisting boundary condition (\ref{bc:y}):
\bea
\notag\expval{\Psi(\Theta_y)|\hat P_x|\Psi(\Theta_y)}=e^{\imth L_y\bar\rho}\expval{\Psi(\Theta_y+\phi L_y)|\hat P_x|\Psi(\Theta_y+\phi L_y)}
\eea
Since polarization is physically the ``center of mass'' of all charges, its change indicates pumping of a charge $L_y\bar\rho$ upon insertion of flux $\phi L_y$ (and hence change of boundary condition). This is a direct evidence for nontrivial Hall conductance $\sigma_{xy}\neq0$ in the gapped ground state. Similar arguments apply to Thm. \ref{thm:b:U(1)xU(1)}-\ref{thm:b:U(1)xZq} as well.

Generally with global symmetry $G_s$, the change of ``generalized polarization'' under translation $\hat T_x$ must be compatible with the pumping of ``generalized symmetry charges'' upon flux insertion (\ie boundary condition twist). This implies a symmetric SRE ground state must be a SPT phase with certain (generalized) flux-charge binding. For instance in fermion symmetry class D (Thm. \ref{thm:f:class D}), we consider ``fermion parity polarization'' which is nothing but the total fermion parity $(-1)^{\hat F}$ on $L_y=$~odd torus: it changes sign by either translation $\hat T_x$. The only compatible SRE ground state is a $\nu=$~odd chiral TSC in class D, such as spinless $p_x+\imth p_y$ superconductor\cite{Read2000}, whose fermion parity changes upon switching between periodic and antiperiodic boundary conditions. In other examples with time reversal symmetry (Thm. \ref{thm:f:class AII}-\ref{thm:b:U(1)+T}) and an odd number of Kramers doublets in each u.c., one can observe the change of ``time reversal polarization''\cite{Fu2006} under translation $\hat T_x$: it is detected by the presence/absence of Kramers pairs in entanglement spectra\cite{Turner2011,Watanabe2015} at different entanglement cuts related by $\hat T_x$ (\eg $\bar x$ and $\bar x+1$ in FIG. \ref{fig:square}). This implies the pumping of one Kramers doublet by switching between periodic and antiperiodic boundary conditions ($\phi L_y=\pi\mod2\pi$), only compatible with a QSH ground state\cite{Fu2006,Levin2009}. All theorems in TABLE \ref{tab:2d:fermion}-\ref{tab:2d:boson} can be proven following this line of thoughts\cite{Supp}.


\begin{figure}[h]
\centering
\includegraphics[width=0.48\columnwidth]{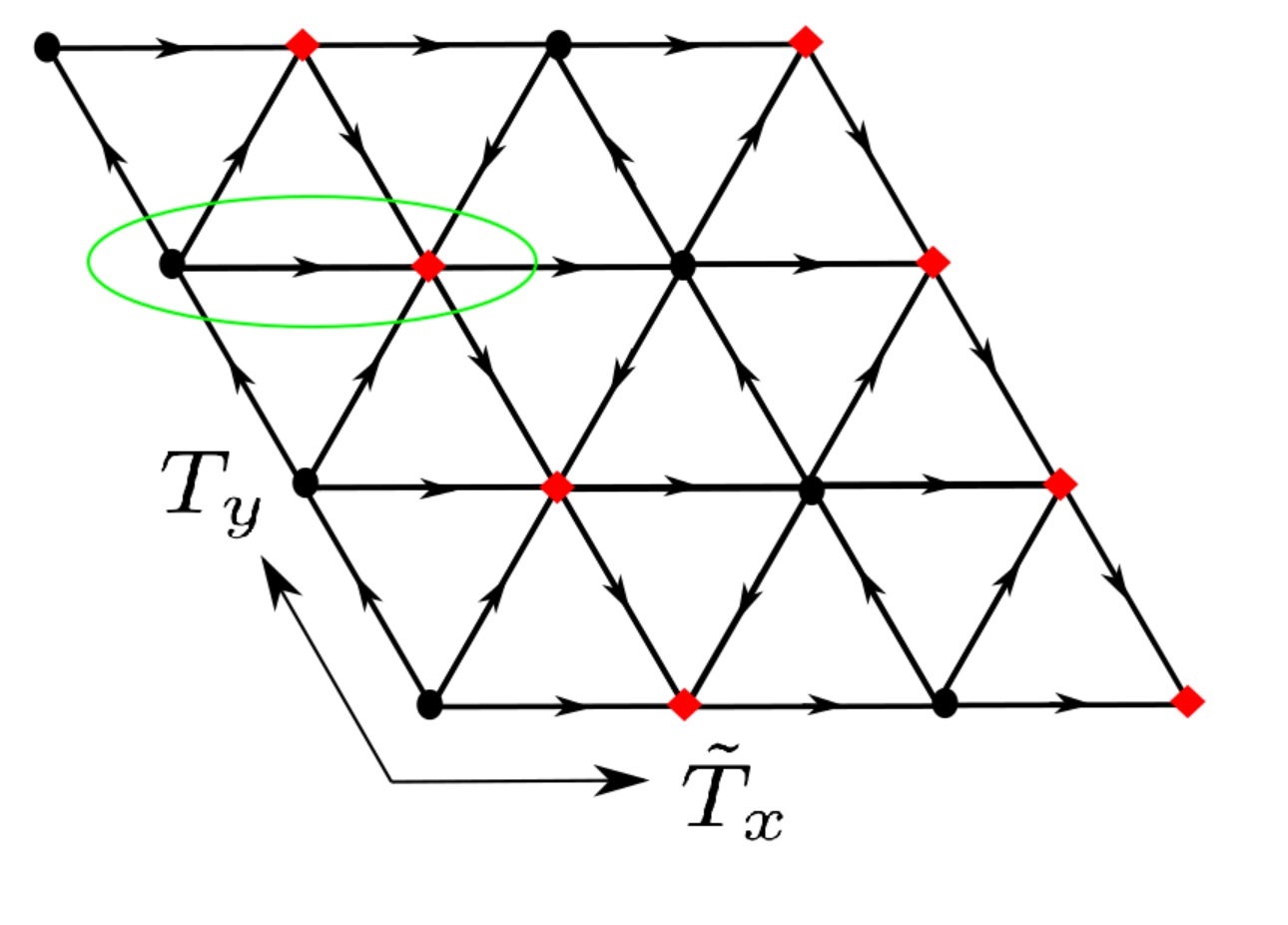}
\includegraphics[width=0.5\columnwidth]{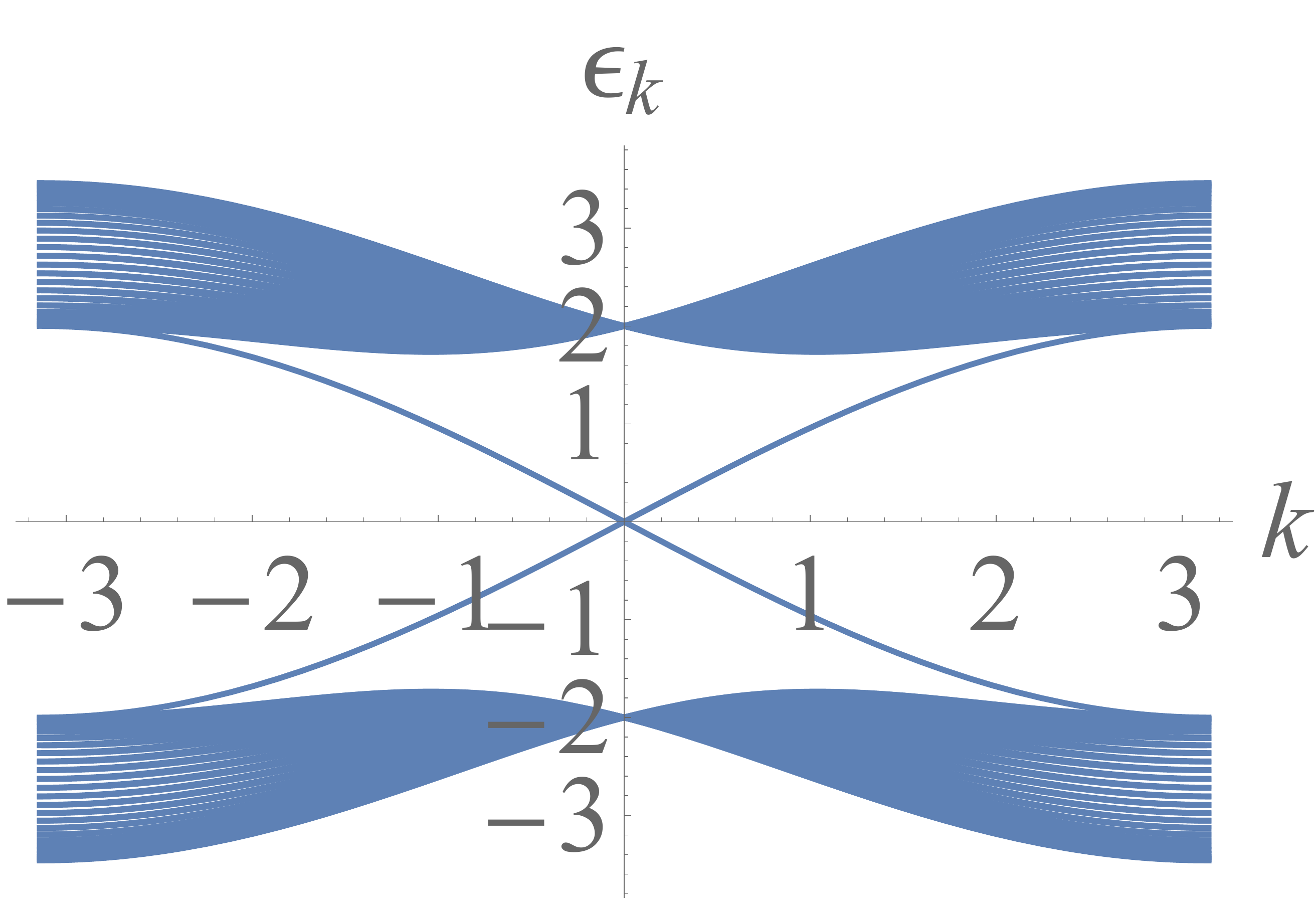}
\caption{(color online) The NN tunneling model of MZMs on a triangular vortex lattice (left) and its edge spectrum (right). The doubled magnetic unit cell is labeled by the green oval. The edge spectrum is obtained on a $L_y=50$ cylinder periodic along $\tilde T_x$ direction, where two counter-propagating edge modes are located separately on two opposite edges.}
\label{fig:vortex lattice}
\end{figure}

\section{Applications}

{\bf Majorana vortex lattice}: The simplest application of Thm. \ref{thm:f:class D} (fermions in symmetry class D) is the vortex lattice of a 2d chiral $p$-wave TSC\cite{Read2000}, or of a 3d TI-superconductor heterostructure\cite{Fu2008}. In both cases there is a single MZM in each vortex core. Magnetic translation symmetry also naturally emerges in a vortex lattice.

As shown in Ref.\cite{Grosfeld2006}, on a Majorana vortex lattice whose plaquette is a polygon of $n$ vortices, there is a $(n\frac{\pi}2-\pi)$ flux per plaquette in a Majorana hopping model between the vortices. Therefore on both triangular ($n=3$) and square ($n=4$) lattice, there is only one Majorana fermion per u.c. whose tunneling amplitudes preserve magnetic translation (\ref{magnetic translation}) with a flux density of $\phi=\pi$. According to Thm. \ref{thm:f:class D}, any unique gapped ground state of the Majoranas must be a $\nu=$~odd TSC with chiral Majorana edge modes. This is precisely the case for a triangular vortex lattice, as shown in FIG. \ref{fig:vortex lattice}.

{\bf Quantum spin liquids}\cite{Savary2017} provide another platform to realize magnetic translation of fermionic spinons with an emergent gauge flux of $\phi=\pi$ per u.c.\cite{Affleck1988a}. On square lattice, Lieb's theorem\cite{Lieb1994} dictates a $\pi$-flux per square plaquette in lowest-energy spinon ground state, in the presence of particle-hole symmetry.

One well-known example is the $U(1)$ Dirac spin liquid in square-lattice large-$N$ Heisenberg model\cite{Affleck1988a}, where $N$-flavor fermionic spinons at half filling ($\bar\rho=\frac12$ for each flavor) hop in the background of $\phi=\pi$ flux per plaquette. According to Thm. \ref{thm:f:class A}, gapping out the Dirac spectrum of the $U(1)$ spin liquid without breaking translation symmetry will result in an odd Chern number $\sigma_{xy}=1\mod2$, corresponding to a chiral spin liquid\cite{Kalmeyer1987,Wen1989} in the context of a spin system.

Another example is in Kitaev-type $Z_2$ spin liquids\cite{Kitaev2006}, where fermionic spinons form a ``superconductor'' and they can see a background $Z_2$ gauge flux of $\phi=0,\pi$. Due to particle-hole symmetry in superconductors, Lieb's theorem\cite{Lieb1994} again applies and points to a $\pi$-flux ground state on square lattice. Ref.\cite{Nakai2012} introduces such a NN square-lattice model, where one Kramers pair of Majorana spinons $\{\gamma_{{\bf r},\uparrow},\gamma_{{\bf r},\downarrow}\}$ per site ${\bf r}$ hops under an emergent $\phi=\pi$ flux as described in (\ref{pi-flux:DIII}). According to Thm. \ref{thm:f:class DIII}, gapping out the Dirac spinon spectrum of NN model (\ref{pi-flux:class D}) while preserving translation symmetry must lead to a TSC of Majorana spinons: as realized by 3-spin interactions involving NNNs\cite{Nakai2012}.

{\bf BIQH states}\cite{Lu2012a,Senthil2013} are also important applications of LSM theorems in TABLE \ref{tab:2d:boson}. In Hofstadter models\cite{Hofstadter1976} of bosons hopping in a magnetic field of $\phi$ flux per u.c., the Hall conductance of a unfractionalized insulator ground state must satisfy relation (\ref{lsm:U(1)}) for one-component bosons, or relation (\ref{boson:U(1)xU(1)}) for two-component bosons. Hence we can use LSM theorems to choose proper boson fillings that enforces a BIQH ground state.

One example is the correlated hopping model of half-filled bosons on honeycomb lattice introduced in Ref.\cite{He2015a}. With only hoppings within the same sublattice, there are two $U(1)$ conservation laws (one on each sublattice) with $\bar\rho_A=\bar\rho_B=\frac12$, and in the presence of $\phi_A=\phi_B=\pi$ flux we must have $\sigma_{xy}^{AB}=1\mod2$ for a unique gapped ground state, as dictated by Thm. \ref{thm:b:U(1)xU(1)}. Indeed this is the SPT state observed in numerical studies of Ref.\cite{He2015a}.

\section{Discussions}

In this work we introduce and prove a new class of LSM theorems in 2d, which relies on magnetic translation symmetries in contrast to the usual LSM theorems with pure lattice translations. While no symmetric SRE ground states is allowed in usual LSM theorems, our theorems imply that a symmetric SRE ground state at fractional filling must belong to a SPT phase with protected edge modes. As summarized in TABLE \ref{tab:2d:fermion}-(\ref{tab:2d:boson}), our LSM theorems apply to many different physical systems with various global symmetries. They will serve as useful guidance to construct realistic models of interacting (especially bosonic) SPT phases, and to future experimental realizations of SPT phases. While the current work focuses on 2d systems with magnetic translation symmetries, it will be interesting to generalize these ideas to other magnetic space group symmetries and to higher spatial dimensions such as 3d, which we leave for future work.

\acknowledgments I am indebted to Ying Ran and Masaki Oshikawa for related collaborations\cite{Lu2017}, Yin-Chen He for discussions on Ref.\cite{He2015a}, and Michael Levin for useful discussions and especially for pointing out Ref.\cite{Grosfeld2006}. I thank KITP for hospitality during ``topoquant16'' program
where part of this work was performed. This work is supported by startup funds at Ohio State University
(YML), and in part by the National Science Foundation under Grant No. NSF PHY11-25915 (YML).


%

\newpage

\appendix
\begin{widetext}
\begin{center}
{{Supplemental Materials}}
\end{center}

\section{Proofs of LSM theorems for free fermions}

\subsection{Symmetry class D}

\subsubsection{Majorana hopping model (\ref{pi-flux:D}) on square lattice}

With only nearest neighbor hoppings on square lattice, the $\pi$-flux model with one Majorana $\gamma_{\bf r}$ per site ${\bf r}$ writes
\bea\label{pi-flux:class D}
&\hat H_0^D=\sum_{\bf r}\imth[t_x\gamma_{\bf r}\gamma_{{\bf r}+\hat x}+t_y(-1)^x\gamma_{\bf r}\gamma_{{\bf r}+\hat y}]+h.c.\\
&\notag=-\sum_{\bf k}\phi_{-{\bf k}}^T\big[2t_y\sin k_y\tau_z+t_x\big(\sin k_x\tau_x+(1-\cos k_x)\tau_y\big)\big]\phi_{\bf k}\notag
\eea
where we define 2-component spinor
\bea
\phi_{\bf k}\equiv\frac1{\sqrt{L_xL_y/2}}\sum_{x,y}e^{-\imth(k_xx+k_yy)}\bpm\gamma_{2x,y}\\ \gamma_{2x+1,y}\epm
\eea
and $\vec\tau$ are Pauli matrices for the sublattice index in a doubled magnetic unit cell (u.c.). Clearly the dispersion vanishes at two Majorana cones (two ``valleys'') at $(k_x,k_y)=(0,0)$ and $(0,\pi)$. A mode expansion around these Majorana cones leads to 4-component spinor
\bea\label{dirac ham:class D}
&\hat H_0^D=-\sum_{\bf q}\Phi_{-\bf q}^T\big[t_x q_x\tau_x+2t_y q_y\tau_z\mu_z\big]\Phi_{\bf q}+O(|{\bf q}|^2),\\
&\Phi_{\bf q}^T\equiv\big(\phi_{\bf q}^T,\phi^T_{(0,\pi)+{\bf q}}\big).\notag
\eea
where $\vec\mu$ are Pauli matrices for the valley index. Under magnetic translations the Majoranas transform as
\bea
\notag&\gamma_{(x,y)}\overset{T_y}\longrightarrow\gamma_{(x,y+1)},\\
&\gamma_{(x,y)}\overset{\tilde T_x}\longrightarrow(-1)^y\gamma_{(x+1,y)}.\label{mag trans:D}
\eea
Therefore the 4-component low-energy spinor transforms as
\bea
&\Phi_{\bf q}\overset{T_y}\longrightarrow\mu_z\Phi_{{\bf q}},\\
&\Phi_{\bf q}\overset{\tilde T_x}\longrightarrow\tau_x\mu_x\Phi_{{\bf q}}.
\eea
It's straightforward to see that among all possible mass terms $\tau_y\mu_{0,x,z}$ and $\tau_z\mu_y$ to Dirac Hamiltonian (\ref{dirac ham:class D}), the only mass that preserves magnetic translation $\{\tilde T_x,T_y\}$ is
\bea\label{dirac mass:class D}
\hat M=m\cdot\hat\Gamma_0,~~~\hat\Gamma_0=\tau_y\mu_z
\eea
This symmetric mass drives the system into a spinless $p_x\pm\imth p_y$ TSC, whose chirality $\nu=\text{Sgn}(m)$ depends on the sign of the mass term. This mass term can be realized by next nearest neighbor (diagonal) hoppings that preserve magnetic translations.\\

\subsubsection{Proof of Theorem \ref{thm:f:class D} for free fermions}

For a generic free-fermion system with an odd number of Majoranas per u.c., one can go beyond perturbing around nearest neighbor model (\ref{pi-flux:class D}), and prove Theorem \ref{thm:f:class D} non-perturbatively. In particular, the momentum-space spinor $\phi_{\bf k}$ transforms under magnetic translations as
\bea
&\phi_{\bf k}\overset{T_y}\longrightarrow e^{\imth k_y}\phi_{\bf k},\\
&\phi_{(k_x,k_y)}\overset{\tilde T_x}\longrightarrow\bpm0&1\\e^{\imth k_x}&0\epm\phi_{(k_x,k_y+\pi)}\label{ff:mts:D}
\eea
Generically a gapped 2d superconductor in class D can be diagonalized as
\bea
&\hat H^D=\sum_{\bf k}\phi^{T}_{-{\bf k}}\hat h_{\bf k}\phi_{\bf k}=\sum_{\bf k}\Gamma_{-{\bf k}}^{T}\hat \Lambda_{\bf k}\Gamma_{\bf k},\\
\notag&\Lambda_{\bf k}=\imth\oplus_{{\bf k}}\bpm0&E_{\bf k}\\-E_{\bf k}&0\epm,~~~E_{\bf k}>0,\\
&\Gamma_{\bf k}=W_{\bf k}\phi_{\bf k},~~~W_{-\bf k}=W_{\bf k}^\ast.
\eea
where $W_{\bf k}$ is a unitary matrix representing the Bloch wavefunction at momentum ${\bf k}$. As shown in \Ref{Kitaev2001,Tewari2012,Budich2013}, the parity of topological index $\nu$ for a 2d superconductor in class D is given by
\bea
&(-1)^\nu=\text{Sgn}\Big(\text{Pf}(\hat h_{k_x=0})\cdot\text{Pf}(\hat h_{k_x=\pi})\Big)=\text{Sgn}\Big(\prod_{{\bf Q}=-{\bf Q}}\det{W_{\bf Q}}\Big)
\eea
where $\text{Pf}(\hat h)$ denotes the Pfaffian of antisymmetric matrix $\hat h$. Here ${\bf Q}=-{\bf Q}$ represents the 4 time reversal invariant momenta (TRIM) \ie $(0,0)$, $(0,\pi)$, $(\pi,0)$ and $(\pi,\pi)$. Meanwhile magnetic translation symmetry (\ref{ff:mts:D}) dictates that
\bea
W_{(k_x,k_y+\pi)}=W_{(k_x,k_y)}\cdot\bpm0&1\\e^{\imth k_x}&0\epm
\eea
and hence we have (notice that $\det W_{\bf Q}=\pm1$ for any TRIM $\bf Q$)
\bea
&\notag\det W_{(0,0)}\cdot\det W_{(0,\pi)}=\det\bpm0&1\\1&0\epm=1,\\
&\notag\det W_{(\pi,0)}\cdot\det W_{(\pi,\pi)}=\det\bpm0&1\\-1&0\epm=-1.
\eea
Therefore we have proven that $\nu$=odd for any gapped superconductor \ie Theorem \ref{thm:f:class D} for any free-fermion system.

\subsection{Symmetry class DIII}

For simplicity, we consider a $\pi$-flux model on square lattice, with an $N_f=$~odd number of Kramers pairs $\{\gamma^a_{{\bf r},\uparrow},\gamma^a_{{\bf r},\downarrow}|1\leq a\leq N_f\}$ per u.c. ${\bf r}$. We can therefore define an odd number of complex fermions per u.c.:
\bea
f_{{\bf r},a}\equiv\frac{\gamma^a_{{\bf r},\uparrow}+\imth\gamma^a_{{\bf r},\downarrow}}2.
\eea
Under time reversal symmetry (TRS) each complex fermion transforms as a Kramers doublet
\bea\label{trs:DIII:0}
f_{{\bf r},a}\overset{\bst}\longrightarrow-\imth f^\dagger_{{\bf r},a},~~~\bst^2=(-1)^{\hat F}=\prod_{{\bf r},a}(\imth\gamma^a_{{\bf r},\uparrow}\gamma^a_{{\bf r},\downarrow}).
\eea
where $\vec\tau$ are Pauli matrices for the Nambu indices. Meanwhile, clearly there is also a particle-hole symmetry (PHS) for the Nambu spinor $\psi_{\bf r}\equiv(f_{\bf r},f^\dagger_{{\bf r}})^T$
\bea\label{sym:DIII:phs}
\psi_{\bf r}=\tau_x\psi_{\bf r}^\ast.
\eea
In this Nambu basis, the BdG Hamilotnian for superconductors is mapped to a Bloch Hamilotnian for band insulators, where the above PHS is essentially a half-filling condition ($N_f$ particles per u.c.) for the ``band insulator''. In the meantime, TRS (\ref{trs:DIII:0}) is implemented as
\bea\label{trs:DIII}
\psi_{\bf r}\overset{\bst}\longrightarrow\tau_y\psi_{\bf r}
\eea

In the presence of magnetic translational symmetry with $\pi$-flux per u.c., a magnetic u.c. consists of 2 u.c. and the Bloch spinor in momentum space is defined as
\bea
\Psi_{{\bf k}=(k_x,k_y)}\equiv\frac1{\sqrt{N/2}}\sum_{x,y}e^{-\imth(xk_x+yk_y)}\bpm\psi_{(2x,y)}\\ \psi_{(2x+1,y)}\epm.
\eea
Under PHS and TRS it transforms as
\bea\label{PHS+TRS:DIII}
\Psi_{\bf k}\overset{\text{PHS}}=\tau_x\Psi_{-{\bf k}}^\ast,~~~\Psi_{\bf k}\overset{\bst}\longrightarrow\tau_y\Psi_{-{\bf k}}
\eea
Meanwhile under magnetic translations in the Landau gauge
\bea\label{mag trans:DIII:0}
\tilde T_x=T_x\cdot(-1)^{\sum_{\bf r}y\sum_a f^\dagger_{{\bf r},a}f_{{\bf r},a}}
\eea
the ``Bloch'' spinor $\Psi_{\bf k}$ transforms as
\bea\label{mag trans:DIII}
\Psi_{\bf k}\overset{\tilde T_x}\longrightarrow\bpm0&1\\e^{\imth k_x}&0\epm_{\vec\mu}\Psi_{(k_x,k_y+\pi)},~~\Psi_{\bf k}\overset{T_y}\longrightarrow e^{\imth k_y}\Psi_{\bf k}.
\eea
Clearly the magnetic translation $\tilde T_x$ shifts the $k_y$ component by $\pi$. We use Pauli matrices $\vec\mu$ for sublattice index, and $\vec\tau$ for the Nambu index.

As shown in Ref.\cite{Fu2006,Budich2013a}, the $Z_2$-valued bulk invariants of topological superconductors in symmetry class DIII is quite similar to the QSHE in symmetry class AII\cite{Fu2006}, given by the ``time reversal polarization''
\bea\label{fu-kane formula}
\nu=(-1)^{P_\Theta(k_y=0)-P_\Theta(k_y=\pi)}\in\pm1
\eea
In particular, the time reversal polarization is given by\cite{Fu2006}
\bea\label{time reversal polarization}
(-1)^{P_\Theta(k_y)}=e^{\frac\imth2(\int_0^\pi\text{d}k_x-\int_{-\pi}^0\text{d}k_x)A^x_{{\bf k}}
}\frac{\text{Pf}[w{(k_x=\pi,k_y)}]}{\text{Pf}[w{(k_x=0,k_y)}]}
\eea
We define the Berry connection for all filled bands
\bea
{\bf A}_{\bf k}=(A^x_{\bf k},A^y_{\bf k})\equiv\imth\sum_{\alpha=\text{filled}}\dbra{{\bf k},\alpha}\vec\nabla_{\bf k}\dket{{\bf k},\alpha}
\eea
and the anti-symmetric ``time reversal'' matrix
\bea
w_{\alpha,\beta}({\bf k})\equiv\langle-{\bf k},\alpha|\bst|{\bf k},\beta\rangle,~~~\alpha,\beta=\text{filled}.
\eea
In our case with magnetic translational symmetry (\ref{mag trans:DIII}), it's straightforward to show that
\bea
&\notag A^x_{(k_x,k_y)}-A^x_{(k_x,k_y+\pi)}=-\sum_{\alpha=\text{filled}}\langle{\bf k},\alpha|\bpm1&0\\0&0\epm_{\vec\mu}\dket{{\bf k},\alpha}\\
&=-\sum_{\alpha=\text{filled}}\dbra{(k_x,k_y+\pi),\alpha}\bpm0&0\\0&1\epm_{\vec\mu}\dket{(k_x,k_y+\pi),\alpha}\notag\\
&
\eea
Note that one can always choose a (smooth) gauge so that time reversal symmetry (\ref{PHS+TRS:DIII}) leads to
\bea
w_{\bf Q}\equiv(\imth\tau_y)\otimes\hat 1_{N_f\times N_f}
\eea
at the four TRIM ${\bf Q}=-{\bf Q}$. Therefore the Pfaffians in (\ref{time reversal polarization}) cancels out each other, and the $Z_2$-valued invariant is simply given by
\bea
&\notag\nu=(-1)^{P_\Theta(k_y=0)-P_\Theta(k_y=\pi)}\\
&=e^{-\frac\imth2\int_0^\pi\text{d}k_x\sum_{\alpha=\text{filled}}
\langle{(k_x,0),\alpha}|(k_x,0),\alpha\rangle}\notag\\
&=e^{-\imth\pi\frac{2N_f}2}=(-1)^{N_f}
\eea
As a result, we have shown that a gapped superconducting ground state can only be a $\nu=-1$ topological superconductor in symmetry class DIII, if we have $N_f=$~odd Kramers pairs of Majorana fermions per u.c..

\subsection{Symmetry class AII and A}

Symmetry class AII corresponds to topological insulators, with a $\mbz_2$ classification in 2d associated with quantum spin Hall effects. Its bulk topological invariant is also given by time reversal polarization\cite{Fu2006} in (\ref{fu-kane formula})-(\ref{time reversal polarization}). In the presence of magnetic translation symmetry (\ref{mag trans:DIII}) with $\pi$-flux per u.c., the bulk invariant can be computed in complete parallel to previous case of class DIII. One can similarly prove that $\nu=(-1)^{\bar\rho_f}$ when there is $\bar\rho_f$ spin-$1/2$ fermions per u.c.. Therefore a gapped ground state can only be a quantum spin Hall insulator with $\nu=-1$, at half-filling with $\bar\rho_f=$~odd.

In the case insulators (symmetry class A) with $\phi$ flux and $\bar\rho_f$ fermions per u.c., the theorem
\bea
\sigma_{xy}\frac{\phi}{2\pi}=\bar\rho_f\mod1.
\eea
was proved in the context of free fermion band theory\cite{Dana1985}. In the special case of $\phi=2\pi/q$ and $\bar\rho_f=p/q$, it reduces to the formula
\bea
\sigma_{xy}=p\mod q
\eea

\section{Proofs of LSM theorems for interacting systems}

Without loss of generality, we always consider a square lattice for simplicity. In the case of square lattice, each u.c. consists of just one site, therefore we also refer it to a site in proper context. We will always choose Landau gauge for simplicity, where translation along $\hat y$ direction is the pure crystal translation $T_y$, while along $\hat x$ direction there is a magnetic translation $\tilde T_x$. Our proofs however do not depend on the lattice geometry, as long as magnetic translation symmetries are preserved.

\subsection{Fermion: Symmetry class D}

Consider a generic interacting system with $N_\gamma=$~odd Majorana fermions $\{\gamma_{{\bf r},a}|1\leq a\leq N_\gamma\}$ per site ${\bf r}$ on a $L_x\times L_y$ torus, where $L_y=$~odd and $L_x=$~even. On such a $L_y=$~odd torus, although translation $T_y$ is still intact, the magnetic translation $\tilde T_x$ in (\ref{mag trans:D}) is in fact broken. As shown in FIG. \ref{fig:square}, in contrast to periodic boundary condition along $\hat x$ direction, the boundary condition along $\hat y$ direction will switch between periodic and antiperiodic in different columns. If we translate the torus along $\hat x$ direction by one u.c., the boundary condition along $\hat y$ direction will be twisted by a phase factor of $e^{\imth\pi}=-1$. In other words, denoting a ground state $\dket{\Psi}$ with boundary condition $e^{\imth\Theta_y}$:
\bea\label{app:bc:y}
&\Psi({\bf X}_1+L_y\hat y,{\bf X}_2,\cdots)=e^{\imth\Theta_y}\Psi({\bf X}_1,{\bf X}_2,\cdots).
\eea
as $\dket{\Psi(\Theta_y)}$, we have
\bea\label{app:twist bc}
\hat T_x\dket{\Psi(\Theta_y)}=\dket{\Psi(\Theta_y+\phi L_y)}.
\eea
where $\phi L_y=\pi\mod2\pi$ here.

Note that with $L_yN_\gamma=$~odd Majorana fermions per column of the torus, the crystal translation $T_x$ plays the role of a supersymmetry\cite{Hsieh2016} which changes fermion parity $(-1)^{\hat F}$:
\bea
T_x (-1)^{\hat F}=(-1)^{\hat F}T_x\cdot(-1)^{L_yN_\gamma}=-(-1)^{\hat F}T_x
\eea
Therefore the two ground states in (\ref{app:twist bc}) related by switching periodic/antiperiodic boundary conditions will have opposite fermion parities if $N_\gamma=$~odd:
\bea
&\notag\expval{\Psi(\Theta_y)|(-1)^{\hat F}|\Psi(\Theta_y)}=\\
&(-1)^{L_y N_\gamma}\expval{\Psi(\Theta_y+\pi L_y)|(-1)^{\hat F}|\Psi(\Theta_y+\pi L_y)}\label{app:fermion parity twist}
\eea

In symmetry class D, all superconductors are classified by an integer index $\nu\in\mbz$. Among them, $\nu=$~odd topological superconductors (\eg spinless $p_x+\imth p_y$ superconductor has $\nu=1$) are distinguished from $\nu=$~even ones with 3 sharp features\cite{Read2000,Kitaev2006,Nayak2008}: (i) an odd number of chiral Majorana modes on an open boundary, with half-integer-valued chiral central charge $c_-=\nu/2$; (ii) one robust Majorana zero mode in each vortex (\ie $\pi$ flux) core; (iii) change of fermion parity if the boundary condition along one direction ($\hat y$-direction in our case) is switched from periodic to anti-periodic.

The last feature (iii) \ie change of fermion parity upon twisting boundary condition along one (say $\hat y$) direction can be intuitively understood as the following, by making a connection to the well-known feature (ii) \ie a single Majorana zero mode trapped at each $\pi$ flux. Twisting boundary condition along $\hat y$ direction is equivalent as dragging a $\pi$ flux across the whole system along $\hat x$ direction. Now that a Majorana bound state is localized around each $\pi$ flux in a $\nu=$~odd topological superconductor, bringing this single Majorana fermion across the system will necessarily change the fermion parity\cite{Kitaev2001,Hsieh2016}.

Therefore in our case with magnetic translation symmetry, as dictated by condition (\ref{app:fermion parity twist}) with $L_y=$~odd, the only unique gapped ground state compatible with feature (iii) must be a $\nu=$~odd topological superconductor. This proves the LSM theorem for class D.

\subsection{Fermion: Symmetry class DIII}

\subsubsection{A no-go theorem for translational symmetric system}

Before proving our LSM theorem with magnetic translation symmetries, we first prove a related theorem for a system with the usual crystal translation symmetry. The no-go theorem states the following:
\begin{theorem}
\label{thm:no-go:DIII}
For a generic interacting fermion system with a $N_f=$~odd number of Kramers pairs of Majoranas $\{\gamma_{{\bf r},a,\sigma}|\sigma=\uparrow/\downarrow,1\leq a\leq N_f\}$ per u.c. ${\bf r}$, there is no unique gapped ground state that preserves both translations and time reversal symmetry.
\end{theorem}

The proof of the theorem is simple. With crystal translational symmetry, we are allowed to put the many-body system on any periodic lattice\cite{Oshikawa2000}, and we choose a $L_x\times L_y$ torus where both lengths are odd:
\bea
L_x,L_y=1\mod2.
\eea
Notice that under time reversal symmetry, the Majorana fermions transform as Kramers doublets
\bea
\bpm\gamma_{{\bf r},a,\uparrow}\\ \gamma_{{\bf r},a,\downarrow}\epm\overset{\bst}\longrightarrow
\bpm\gamma_{{\bf r},a,\downarrow}\\-\gamma_{{\bf r},a,\uparrow}\epm
\eea
and therefore
\bea
\bst(\imth\gamma_{{\bf r},a,\uparrow}\gamma_{{\bf r},a,\downarrow})\bst^{-1}=-\imth\gamma_{{\bf r},a,\uparrow}\gamma_{{\bf r},a,\downarrow}
\eea
Note that the total fermion parity is given by
\bea
(-1)^{\hat F}=\prod_{{\bf r},a}(\imth\gamma_{{\bf r},a,\uparrow}\gamma_{{\bf r},a,\downarrow})
\eea
Therefore the fermion parity and time reversal symmetry satisfy the following algebra
\bea
\bst(-1)^{\hat F}\bst^{-1}(-1)^{\hat F}=(-1)^{N_fL_xL_y}
\eea
This means on a odd by odd torus, time reversal symmetry $\bst$ serves as a supersymmetry\cite{Qi2009} that changes the fermion parity. Since time reversal and fermion parity anticommutes with each other, they cannot be both preserved in a unique symmetric ground state. Therefore we have proved the no-go theorem.

\subsubsection{Proof of LSM theorem \ref{thm:f:class DIII} for class DIII}\label{app:subsec:DIII}

Now let's turn to the case with magnetic translation symmetry (\ref{mag trans:DIII:0}). This time we consider an infinite cylinder which is finite along $\hat y$ direction, but infinite along $\hat x$ direction. Again we choose the circumference length $L_y$ along $\hat y$ direction to be odd. Quite similar to the class D case, the boundary condition along $\hat y$ direction switches between periodic and antiperiodic in different columns of the cylinder. When we translate the physical system by one u.c. along $\hat x$ direction, we twist the $\hat y$-direction boundary condition by a phase of $e^{\imth\pi}=-1$.

To prove the LSM theorem, we first assume a unique gapped ground state that preserves magnetic translation and time reversal symmetries. Using symmetry properties of the entanglement spectrum of a SRE state\cite{Pollmann2010,Watanabe2015}, we are able to show that this SRE ground state must be a topological superconductor in class DIII.

Next we consider the Schmidt decomposition of unique SRE ground state $\dket{\Psi(\Theta_y)}$ with boundary condition (\ref{app:bc:y}) across an entanglement cut along $\hat y$ direction located at $x_0-1<\bar x<x_0$ (see FIG. \ref{fig:square}):
\bea\label{sdcp:before:x}
\dket{\Psi(\Theta_y)}=\sum_{\alpha}\lambda^{\Theta_y}_{\bar x,\alpha}\dket{\alpha,\Theta_y}_{\bar x,L}\dket{\alpha,\Theta_y}_{\bar x,R}
\eea
where $\lambda_{\bar x,\alpha}$ are Schmidt weights. Note that in a generic Hamiltonian of Majorana fermions, the fermion number is not conserved and the Schmidt eigenstates do not generally have a fixed particle number. In the presence of time reversal symmetry $\bst$, although fermion parity $(-1)^{\hat F}$ can fluctuate for each Schmidt state, $\{\dket{\alpha}_{\bar x,L/R}\}$ must form a representation of the following algebra
\bea\label{app:DIII:character:before}
\bst(-1)^{\hat F}\bst^{-1}(-1)^{\hat F}\dket{\alpha,\Theta_y}_{\bar x,L}=e^{\imth\Phi_{\bar x}}\dket{\alpha,\Theta_y}_{\bar x,L}
\eea
where $e^{\imth\Phi_{\bar x}}=\pm1$ is a phase factor depending on the entanglement cut (at $\bar x$), but independent of Schmidt eigenstate $\dket{\alpha}_{\bar x,L}$. Similar to the class D case, the many-body symmetry (\ref{app:twist bc}) for ground state boundary condition exists in class DIII as well. According to relation (\ref{app:twist bc}), a Schmidt decomposition (\ref{sdcp:before:x}) of $\dket{\Psi(\Theta_y)}$ at entanglement cut $\bar x$ leads to the same entanglement spectrum as that of $\dket{\Psi(\Theta_y+\pi L_y)}$ at entanglement cut $\bar x+1$ (see FIG. \ref{fig:square})
\bea
\dket{\Psi(\Theta_y+\pi L_y)}=\sum_{\beta}\lambda^{\Theta_y+\pi L_y}_{\bar x+1,\alpha}\dket{\alpha,\Theta_y+\pi L_y}_{\bar x,L}\dket{\alpha,\Theta_y+\pi L_y}_{\bar x,R}\label{sdcp:after:x+1}
\eea
with
\bea
\lambda^{\Theta_y+\pi L_y}_{\bar x+1,\alpha}=\lambda^{\Theta_y}_{\bar x,\alpha},\\
\dket{\alpha,\Theta_y+\pi L_y}_{\bar x+1,L/R}=\hat T_x\dket{\alpha,\Theta_y}_{\bar x,L}.\label{app:DIII:translation}
\eea
In comparison, the original ground state $\dket{\Psi(\Theta_y)}$ has the following Schmidt decomposition at entanglement cut $\bar x+1$:
\bea\label{sdcp:before:x+1}
\dket{\Psi(\Theta_y)}=\sum_{\beta}\lambda^{\Theta_y}_{\bar x+1,\beta}\dket{\beta,\Theta_y}_{\bar x+1,L}\dket{\beta,\Theta_y}_{\bar x+1,R}
\eea

In the following, we will compare the Schmidt decompositions (\ref{sdcp:before:x+1}) and (\ref{sdcp:after:x+1}) for two ground states with two boundary conditions differed by $\pi L_y=\pi\mod2\pi$, at the same entanglement cut $\bar x+1$. First according to (\ref{app:DIII:character:before}) and (\ref{app:DIII:translation}), clearly Schmidt eigenstates of $\dket{\Psi(\Theta_y+\pi L_y)}$ has a symmetry character $e^{\imth\Phi_{\bar x}}$:
\bea
\bst(-1)^{\hat F}\bst^{-1}(-1)^{\hat F}\dket{\alpha,\Theta_y+\pi L_y}_{\bar x+1,L}=e^{\imth\Phi_{\bar x}}\dket{\alpha,\Theta_y+\pi L_y}_{\bar x+1,L},~~~\forall~\alpha.\label{sdcp:sym:after:x+1}
\eea
What about Schmidt eigenstates $\dket{\beta,\Theta_y}_{\bar x+1,L}$ of ground state $\dket{\Psi(\Theta_y)}$? Note that the Schmidt eigenstates of $\dket{\Psi(\Theta_y)}$ at the two different cuts (\ref{sdcp:before:x}) and (\ref{sdcp:before:x+1}) are related by
\bea\label{x->x+1}
\dket{\beta,\Theta_y}_{\bar x+1,L}=\sum_{p,\alpha}M^p_{\beta,\alpha}\dket{p}_{x_0}\otimes\dket{\alpha,\Theta_y}_{\bar x,L}
\eea
where $\{\dket{p}_{x_0}\}$ is a set of orthonormal basis for Hilbert space on column $x_0$. Note that we have a $L_y=$~odd number of Majorana Kramers pairs on column $x_0$, and therefore
\bea
\bst(-1)^{\hat F}\bst^{-1}(-1)^{\hat F}\dket{p}_{x_0}=(-1)\cdot\dket{p}_{x_0}.
\eea
As a result, the symmetry character of Schmidt eigenstates (\ref{sdcp:before:x+1}) at cut $\bar x+1$ have a symmetry character
\bea
\bst(-1)^{\hat F}\bst^{-1}(-1)^{\hat F}\dket{\beta,\Theta_y}_{\bar x+1,L}=e^{\imth\Phi_{\bar x+1}}\dket{\beta,\Theta_y}_{\bar x+1,L}=e^{\imth(\Phi_{\bar x}+\pi)}\dket{\beta,\Theta_y}_{\bar x+1,L}\label{sdcp:sym:before:x+1}
\eea

Comparing (\ref{sdcp:sym:before:x+1}) and (\ref{sdcp:sym:after:x+1}), we can see that after twisting boundary condition along $\hat y$ direction, for the same entanglement cut at $\bar x+1$, the entanglement spectrum of a ground state $\dket{\Psi(\Theta_y+\pi L_y)}$ with twisted boundary condition has different symmetry character $\bst(-1)^{\hat F}\bst^{-1}(-1)^{\hat F}=e^{\imth\Phi_{\bar x}}$, as compared to the original ground state $\dket{\Psi(\Theta_y)}$ whose entanglement spectrum has symmetry character $\bst(-1)^{\hat F}\bst^{-1}(-1)^{\hat F}=e^{\imth\Phi_{\bar x+1}}=-e^{\imth\Phi_{\bar x}}$.

What kind of symmetric SRE ground state is compatible with the change of Schmidt eigenstate symmetry character upon twisting boundary condition? As discussed in the case of class D, twisting the boundary condition along $\hat y$ direction can be viewed as dragging a $\pi$ flux across the cylinder along $\hat x$ direction. In symmetry class DIII, there are two classes ($\nu\in\mbz_2$ classification) of time-reversal-symmetric (TRS) SRE superconductors: the trivial one with $\nu=0$ and the topological superconductor with $\nu=1$. While the $\pi$ flux in a trivial superconductor has no stalbe low-energy bound states, the $\pi$ flux in a 2d topological superconductor features a zero-energy Majorana Kramers pair\cite{Qi2009} $\{\gamma_\uparrow,\gamma_\downarrow\}$. Therefore, dragging a $\pi$-flux across the cylinder along $\hat x$ direction will also move this Kramers pair of Majoranas across the entanglement cut (at $\bar x+1$). Now that $\bst(-1)^{\hat F}\bst^{-1}(-1)^{\hat F}=-1$ when acting on an odd number of Majorana Kramers pairs, the symmetry character of entanglement spectrum will be switched by twisting boundary condition in such a topological superconductor. Therefore in the presence of magnetic translation with $\pi$ flux per u.c., the only TRS SRE compatible with the above entanglement spectrum symmetry character is the $\nu=1$ topological superconductor. This proves the LSM theorem for symmetry class DIII. \\

\subsection{Symmetry group $G_s=U(1)_A\times U(1)_B$}

In the presence of a global $U(1)$ symmetry, an insulating ground state $\dket{\Psi}$ on a $L_x\times L_y$ torus can be characterized by its polarization $\hat P_x\equiv e^{\frac{2\pi}{L_x}\sum_{\bf r}x\hat n_{\bf r}}$ (and $\hat P_y$ can be defined similarly), where $\hat n_{\bf r}$ is the $U(1)$ charge on lattice site ${\bf r}$\cite{Vanderbilt1993,RestaSorella1999,RestaJPCMrev2002}. Physically the polarization describes the center of mass of all $U(1)$ charges. In contrast to metals with no well-defined polarization \ie $\expval{\Psi|\hat P_\alpha|\Psi}=0$, insulators generally have a non-vanishing complex expectation value of each polarization component $\hat P_x$ and $\hat P_y$.

Similar to previous cases, the boundary condition along $\hat y$ direction generally changes on different columns of the torus. As a result, pure lattice translation $\hat T_x$ can twist the $\hat y$-direction boundary condition as shown in \ref{app:twist bc}. Meanwhile if there is a unique symmetric ground state separated from excitation states by a finite energy gap, one can adiabatically insert flux through the hole along $\hat x$ direction without closing the gap, and $\hat y$-direction boundary condition can be adiabatically twisted in this flux insertion process:
\bea\label{app:flux insertion}
\hat {{\mathcal{F}}_y}(\phi L_y)\dket{\Psi(\Theta_y)}=e^{\imth\phi_0}\dket{\Psi(\Theta_y+\phi L_y)}.
\eea
where $\phi_0$ is an unimportant phase factor, and $\mathcal{F}_y(\phi L_y)$ is the adiabatic $\phi L_y$-flux insertion operator.

Therefore according to (\ref{app:twist bc}) and (\ref{app:flux insertion}), the system on a $L_x\times L_y$ torus has an emergent symmetry in the unique gapped ground state:
\bea\label{twist+translation}
T_x^\prime=\mathcal{F}_y^{-1}(\phi L_y)\cdot T_x
\eea
if we choose Landau gauge (\ie preserving lattice translation $\tilde T_y\equiv T_y$) for the magnetic translation algebra (\ref{magnetic translation:general}) and (\ref{magnetic translation:two}). Now that the insulator ground state has a non-vanishing expectation value for polarization $\hat P_x$, the emergent symmetry operation (\ref{twist+translation}) must preserve the polarization operator $\hat P_x$\cite{Lu2017}.

In the presence of two $U(1)$ charge conservation symmetries $G_s=U(1)_A\times U(1)_B$, we have a more complicated magnetic translation algebra (\ref{magnetic translation:two}). There are also two polarization operators $\hat P^A_x=e^{\frac{2\pi}{L_x}\sum_{\bf r}x\hat n^A_{\bf r}}$ and $\hat P^B_x=e^{\frac{2\pi}{L_x}\sum_{\bf r}x\hat n^B_{\bf r}}$, and the emergent symmetry (\ref{twist+translation}) on $L_x\times L_y$ torus becomes the following:
\bea\label{twist+translation:U(1)xU(1)}
T_x^\prime=\big[\mathcal{F}_y^A(\phi_A L_y)\mathcal{F}_y^B(\phi_B L_y)\big]^{-1}\cdot T_x
\eea
Therefore both polarizations $P_x^A$ and $P_x^B$ must be preserved by the above emergent symmetry operation. Making use of the following commutation relations:
\bea
&T_xP_x^\alpha T_x^{-1}=e^{-2\pi\imth\bar\rho_\alpha L_y}P_x^\alpha,~~~\alpha,\beta=A,B;\\
&\mathcal{F}_y^\alpha(\phi_\alpha L_y)P_x^\beta\big[\mathcal{F}_y^\alpha(\phi_\alpha L_y)\big]^{-1}=e^{\imth\sigma_{xy}^{\beta\alpha}\phi_\alpha L_y}P_x^\beta.
\eea
and by requiring $[P_x^\alpha,T_x^\prime]=0$ we can immediately obtain relation (\ref{boson:U(1)xU(1)}). Therefore we have proved Thm. \ref{thm:b:U(1)xU(1)}. \\

Clearly, $G_s=U(1)$ is a special case of the above discussions, and one can easily prove relation (\ref{lsm:U(1)}) and Thm. \ref{thm:f:class A} by requiring $T^\prime_x$ in (\ref{twist+translation}) commutes with polarization $P_x$. This is studied in detail in Ref.\cite{Lu2017}. Notice that in all discussions, we have not involve the statistics of microscopic particles, therefore the conclusions apply to interacting bosons and/or fermions.

\subsection{Symmetry group $G_s=U(1)_A\times (Z_q)_B$}

As mentioned in main text, breaking $U(1)_A\times U(1)_B$ down to its subgroup $G_s=U(1)_A\times (Z_q)_B$ leads to a $2\mbz\times(\mbz_q)^2$ classification of 2d SPT phases. They are characterized by Hall conductance $\sigma_{xy}^A$ of conserved $U(1)_A$ charges, a $\nu^B\in\mbz_q$ invariant associated with 2d $(Z_q)_B$-SPT phases, plus another $\nu^{AB}\in\mbz_q$ invariant describing the binding of $U(1)_A$ charges to $(Z_q)_B$ flux. Specifically, each $2\pi$ flux of $(Z_q)_B$ symmetry will trap $\nu^{AB}\in\mbz_q$ units of $U(1)_A$ charges.

Again on $L_x\times L_y$ torus under Landau gauge (preserving $T_y$ translation), the ground state $\dket{\Psi(\Theta_y^A,\Theta_y^B)}$ satisfy
\bea\label{app:twist bc:U(1)xZq}
\hat T_x\dket{\Psi(\Theta_y^A,\Theta_y^B)}=\dket{\Psi(\Theta_y^A+\phi_A L_y,\Theta_y^B+\phi_B L_y)}.
\eea
The polarization $P_x^A$ has the following dependence on $\hat y$-direction boundary conditions:
\bea
&e^{-\imth L_y\bar\rho_A}\cdot\expval{\Psi(\Theta_y^A,\Theta_y^B)|\hat P_x^A|\Psi(\Theta_y^A,\Theta_y^B)}=\\
&\notag\expval{\Psi(\Theta_y^A+\phi_A L_y,\Theta_y^B+\phi_B L_y)|\hat P_x^A|\Psi(\Theta_y^A+\phi_A L_y,\Theta_y^B+\phi_B L_y)}
\eea
Now that twisting $\hat y$-direction boundary condition can be achieved by flux insertions through the hole along $\hat x$-direction, we immediately reach the relation (\ref{boson:U(1)xZq}). Therefore we have proven Thm. \ref{thm:b:U(1)xZq}.

\subsection{Symmetry group $G_s=U(1)\rtimes Z_2^\bst$ and $G_s=Z_2\times Z_2^\bst$}

We first consider a boson system with symmetry group $G_s=U(1)\rtimes Z_2^\bst$, consisting of two parts: (i) integer-spin $U(1)$-charged bosons transformed as Kramers singlest ($\bst^2=+1$), and (ii) charge-neutral half-integer spin moments ($\bst^2=-1$). As stated in Thm. \ref{thm:b:U(1)+T}, we consider an odd number of half-integer spins together with $\phi=\pi$ flux of $U(1)$ symmetry in each unit cell. On a circumference-$L_y$ cylinder which is infinitely long along $\hat x$ direction and wrapped around along $\hat y$-direction, we consider a unique gapped ground state $\dket{\Psi(\Theta_y)}$ where $\Theta_y\in[0,2\pi)$ denotes the boundary condition (\ref{bc:y}) along $\hat y$ direction. Again the relation (\ref{app:twist bc}) holds on this infinite cylinder, imposing strong constraints on the ground state properties.

Similar to the proofs of Thm. \ref{thm:f:class DIII}, we again consider the Schmidt decompositions of two states $\dket{\Psi(\Theta_y)}$ and $\dket{\Psi(\Theta_y+\pi L_y)}$ at the same entanglement cut $\bar x+1$. General relations (\ref{sdcp:before:x}) and (\ref{sdcp:after:x+1})-(\ref{sdcp:before:x+1}) still holds in this case, while the symmetry characters of the Schmidt eigenstates in this case becomes:
\bea
&\bst^2\dket{\alpha,\Theta_y}_{\bar x,L}=e^{\imth\Phi_{\bar x}}\dket{\alpha,\Theta_y}_{\bar x,L},~~~e^{\imth\Phi_{\bar x}}=\pm1,~~~\forall~\alpha,\\
&\bst^2\dket{\alpha,\Theta_y+\pi L_y}_{\bar x+1,L}=e^{\imth\Phi_{\bar x}}\dket{\alpha,\Theta_y+\pi L_y}_{\bar x+1,L},~~~e^{\imth\Phi_{\bar x}}=\pm1,~~~\forall~\alpha,\\
&\bst^2\dket{\beta,\Theta_y}_{\bar x+1,L}=e^{\imth\Phi_{\bar x+1}}\dket{\beta,\Theta_y+\pi L_y}_{\bar x+1,L},~~~e^{\imth\Phi_{\bar x+1}}=\pm1,~~~\forall~\beta.
\eea
Again due to relation (\ref{x->x+1}) and the fact
\bea
\bst^2\dket{p}_{x_0}=(-1)^{L_y}\dket{p}_{x_0},~~~\forall~p.
\eea
we can easily show that
\bea
e^{\imth\Phi_{\bar x+1}}=e^{\imth(\Phi_{\bar x}+\pi)},~~~\text{if}~~L_y=1\mod 2.
\eea
Therefore on a $L_y=$~odd cylinder, the $\bst^2=\pm1$ symmetry character of Schmidt eigenstates of ground state $\dket{\Psi(\Theta_y)}$ at entanglement cut at $\bar x+1$ changes sign when boundary condition $\Theta_y$ is changed by $\pi$. This indicates the pumping of one Kramers doublet across the cylinder as induced by the $\pi$ flux insertion, which is only compatible with a BQSH state where $\pi$ flux is bound to a Kramers doublet. Therefore we have prove Thm. \ref{thm:b:U(1)+T}.

Clearly the above proof goes through even if $U(1)$ symmetry is broken down to a discrete subgroup $H\in U(1)$, as long as $H$ contains $Z_2$ as a subgroup. Therefore the LSM theorem for $G_s=Z_2\times Z_2^\bst$ is also proven.

The proof of Thm. \ref{thm:f:class AII} is very similar to the above discussions, also making use of Schmidt decomposition and symmetry characters of the Schmidt eigenstates. It is discussed in details by Ref.\cite{Wu2017}.\\

\end{widetext}

\end{document}